\documentclass{iau}
\usepackage{graphicx,natbib}
 
\newcommand{\apj}{ApJ}           
\newcommand{\mnras}{MNRAS}       

\newcommand{\aap}{A\&A}

\newcommand{\aj}{AJ}
\newcommand{\pasp}{PASP}

\usepackage[T1]{fontenc}
\newcommand{\changefont}[3]{\fontfamily{#1}\fontseries{#2}\fontshape{#3}\selectfont}
\newfont{\vcap}{cmssdc10 scaled 1000}
\newfont{\lcap}{cmssdc10 scaled 1100}
\newfont{\nlx}{cmssdc10 scaled 900}
\newcommand{\rem}[1]{\nlx {#1}\normalfont}
\newcommand{\trem}[1]{\xnlx {#1}\normalfont}
\newfont{\xnlx}{cmssdc10 scaled 800}
\newfont{\hvss}{cmssdc10 scaled 1540}

\def\rr{{\sl R}$^{\star}$}

\def\arcsec{arcsec}
\def\lyc{{\sl Ly}$_{\rm c}$}

\def\ige{\changefont{cmtt}{m}{it}ne\rm}
\def\wim{\changefont{cmtt}{m}{it}wim\rm}
\def\wism{\wim}

\def\ha{H$\alpha$}
\def\ewha{EW(\ha)}

\def\lo3hb{$\log$([O\,{\sc iii}]/H$\beta$)}
\def\ln2ha{$\log$([N\,{\sc ii}]/H$\alpha$)}
\def\tauha{$\tau$}
\def\tauha_ext{$\tau$}

\def\sisp{\nlx sisp\rm} 
 
\def\la{\lesssim}

\def\sisp{\nlx sisp\rm} 
\def\isan{\nlx isan\rm} 

\def\lfrac{${\cal L}_{\rm 5\,Gyr}$}

\def\m5{${\cal M}_{\star,{\rm 5\,Gyr}}$}

\def\uflux{erg\,s$^{-1}$\,cm$^{-2}$}
\newcommand{\kmsec}{km~s$^{-1}$}

\def\tauha{$\tau$}
\def\tauha_ext{$\tau$}

\def\n2ha{[N\,{\sc ii}]/H$\alpha$}
\def\ln2ha{$\log$([N\,{\sc ii}]/H$\alpha$)}
\def\tn2ha{[N\,{\sc ii}]${\scriptstyle 6583}$/H$\alpha$}
\def\tln2ha{$\log$([N\,{\sc ii}]${\scriptstyle 6583}$/H$\alpha$)}
\def\o3hb{[O\,{\sc iii}\,]/H$\beta$}
\def\lo3hb{$\log$([O\,{\sc iii}]/H$\beta$)}
\def\to3hb{[O\,{\sc iii}]${\scriptstyle 5007}$/H$\beta$}
\def\tlo3hb{$\log$([O\,{\sc iii}]${\scriptstyle 5007}$/H$\beta$)}
\newcommand{\PutLabel}[3]{\put(#1,#2){#3}}
\newcommand {\aga} {\ {\raise-.5ex\hbox{$\buildrel>\over\sim$}}\ }
\newcommand {\ala} {\ {\raise-.5ex\hbox{$\buildrel<\over\sim$}}\ } 
\title{Extended nebular emission in CALIFA early-type galaxies}

\author[Gomes, Papaderos, Kehrig, V\'{\i}lchez, Lehnert and the CALIFA collaboration]
{J.M. Gomes$^1$, P. Papaderos$^1$, C. Kehrig$^2$, J.M. V\'{\i}lchez$^2$, M.D. Lehnert$^3$  and the CALIFA collaboration}

\affiliation{$^1$Instituto de Astrof\'isica e Ci\^encias do Espa\c{c}o, Universidade do Porto, 
CAUP, Rua das Estrelas, PT4150-762 Porto, Portugal\\
email: {\tt jean@astro.up.pt; papaderos@astro.up.pt} \\
$^2$Instituto de Astrof\'isica de Andaluc\'ia (CSIC), Glorieta de la Astronom\'{\i}a s/n Aptdo. 3004, E18080-Granada, Spain\\
$^3$Institut  d'Astrophysique  de  Paris,  UMR  7095,  CNRS,  Universit\'e Pierre et Marie Curie, 98 bis boulevard Arago, 75014 Paris, France
}

\pubyear{2014}
\volume{309}
\jname{Galaxies in 3D across the Universe}
\editors{B. L. Ziegler, F. Combes, H. Dannerbauer, M. Verdugo, eds.}

\begin{document}

\maketitle

\begin{abstract}
The morphological, spectroscopic and kinematical properties of the warm interstellar medium (\wim) in early-type galaxies (ETGs) 
hold key observational constraints to nuclear activity and the buildup history of these massive quiescent systems.
High-quality integral field spectroscopy (IFS) data with a wide spectral and spatial coverage, such as those 
from the CALIFA survey, offer a precious opportunity for advancing our understanding in this respect.
We use deep IFS data from CALIFA ({\tt califa.caha.es}) to study the \wim\ over the entire extent and optical spectral 
range of 32 nearby ETGs.
We find that all ETGs in our sample show faint (\ha\ equivalent width \ewha$\sim$0.5 \dots 2 \AA) extranuclear nebular emission 
extending out to $\geq$2 Petrosian$_{50}$ radii.
Confirming and strengthening our conclusions in Papaderos et al. (2013; hereafter P13) we argue that ETGs span a 
broad \emph{continuous} sequence with regard to the properties of their \wim, and they can be roughly subdivided into two 
characteristic classes. 
The first one (type~i) comprises ETGs with a nearly constant \ewha$\sim$1--3 \AA\ in their extranuclear component, 
in quantitative agreement with (even though, no proof for) the hypothesis of photoionization by the post-AGB 
stellar component being the main driver of extended \wim\ emission. 
The second class (type~ii) consists of virtually \wim-evacuated ETGs with a large
Lyman continuum (\lyc) photon escape fraction and a very low ($\leq$0.5~\AA) \ewha\ in their 
nuclear zone. These two ETG classes appear indistinguishable from one another by their 
LINER-specific emission-line ratios.
Additionally, here we extend the classification by P13 by the class~i+ which stands 
for a subset of type~i ETGs with low-level star-forming activity in contiguous spiral-arm like features 
in their outermost periphery.
These faint features, together with traces of localized star formation in several 
type~i\&i+ systems point to a non-negligible contribution from young massive stars to the global 
ionizing photon budget in ETGs. 
Moreover, our results further highlight the considerable diversity of ETGs
with respect to their gaseous and stellar kinematics. 
Whereas in one half of our sample gas and stars show similar (yet no identical) 
velocity patterns, both dominated by rotation along the major galaxy axis, our CALIFA 
data also document several cases of kinematical decoupling or stellar rotation along 
the minor galaxy axis.
\end{abstract}

\firstsection
\section{Introduction}
Even though the presence of faint nebular emission (\ige) in the nuclei of many early-type galaxies (ETGs) 
has long been established observationally \citep[e.g.,][]{Sarzi10,K12},
the nature of the dominant excitation mechanism of the warm interstellar medium (\wism)
in these systems remains a subject of debate.
The {\sl low-ionization nuclear emission-line region} (LINER) emission-line ratios,
as a typical property of ETG nuclei, have prompted various interpretations
\citep[see, e.g.,][]{YanBlanton2012},
including low-accretion rate active galactic nuclei \citep[AGN; e.g.,][]{Ho2008}, 
fast shocks \citep[e.g.][]{DopitaSutherland1995}, and hot, evolved ($\geq 10^8$ yr) post-AGB (pAGB) 
stars \citep[e.g.,][]{bin94,sta08}.
High-quality integral field spectroscopy (IFS) data with a wide spectral and spatial coverage, 
such as those from the {\sl Calar Alto Legacy Integral Field Area} (CALIFA) survey 
\citep[][]{Sanchez2012}, offer an important opportunity to gain insight into the nature 
of nuclear and extranuclear gas excitation sources and advance our understanding on the 
evolutionary pathways of ETGs.
%
\begin{figure}
\begin{picture}(14.0,3.0)
\put(3.0,0.0){\includegraphics[width=6.4cm, clip=true, viewport=20 110 750 440]{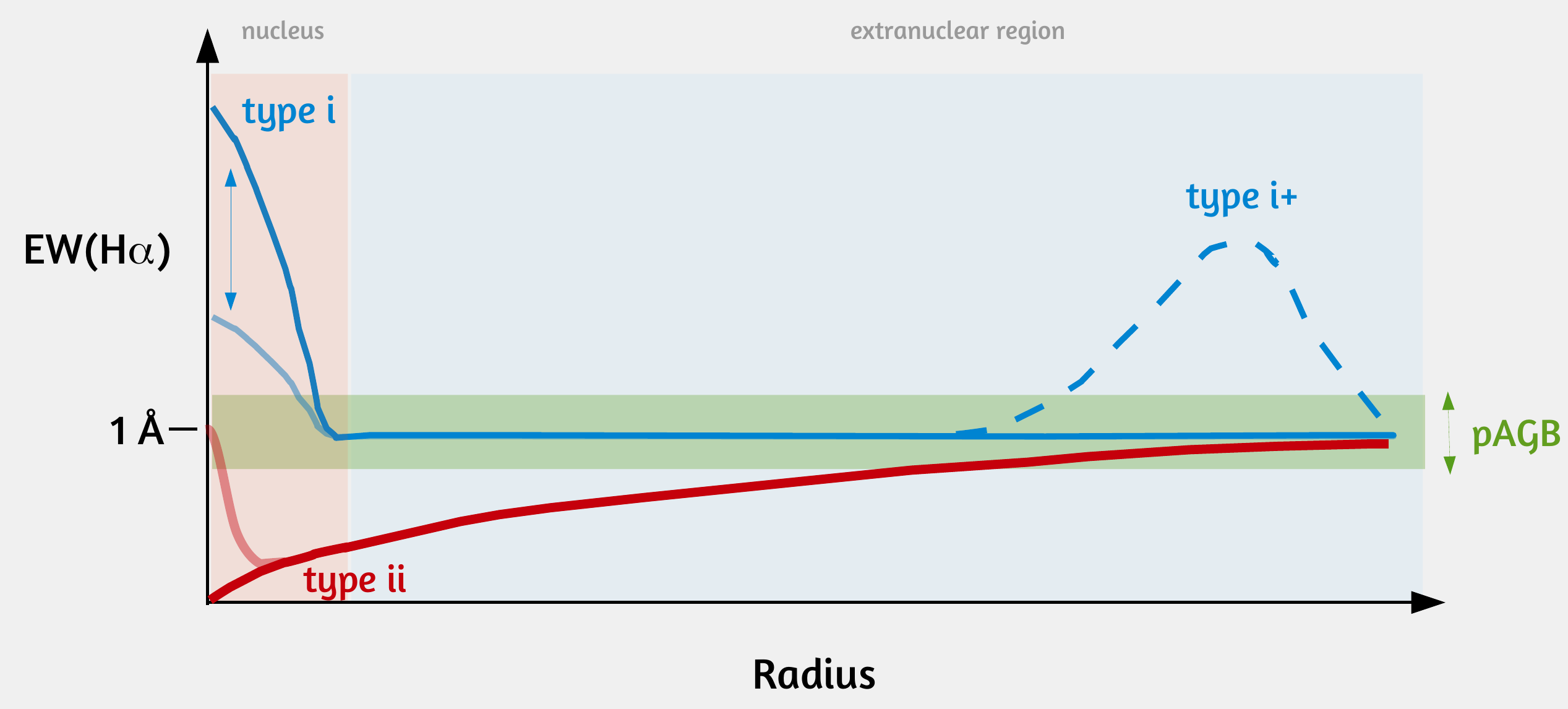}}
\end{picture}
\caption[]{Schematic representation of the two main classes of ETGs, as
  defined in P13 and G14. The \ewha\ profiles of \trem{type~i} ETGs show in their
  extranuclear component nearly constant values within the narrow range
  between $\sim$0.5 \AA\ and $\sim$2.4 \AA, with a mean value of typically $\sim$1~\AA.
  A few of these systems (labeled i+) additionally show an \ewha\ excess in their periphery, 
  which, as discussed in G14, is due to low-level star-forming activity.
The defining property of \trem{type~ii} ETGs is a centrally very low
  ($\leq$0.5~\AA) mean \ewha, increasing then smoothly to $\sim$1~\AA\ at
  their periphery. Regarding their nuclear properties, both ETG types show a
  large diversity, from systems with virtually \wim-evacuated cores
  ($\la$0.1~\AA) to galaxies with a compact nuclear \ewha\ excess.}
\end{figure}
Here we provide a brief summary of our results from an ongoing study of low-spectral-resolution ($R\sim 850$) 
CALIFA IFS cubes, observed with PMAS/PPAK \citep{Roth05,Kelz06}, for 20 E and 12 S0 nearby ($<$150~Mpc) galaxies.
This sample was initially studied in Papaderos et al. (2013; P13) with main focus on the radial distribution of 
the \ewha\ and Lyman continuum photon escape fraction (\lyc) in ETGs, and has permitted a tentative subdivision 
of these systems into two main classes. 
A thorough 2D analysis of the same sample, including, e.g., \ewha\ and stellar age maps, 
stellar and gas kinematics and gas excitation diagnostics will be presented in Gomes et al. (2014; hereafter G14). 
G14 also provide a detailed description of our IFS data processing and spectral modeling pipeline \trem{Porto3D} 
and of the methods used to determine the radial distribution of various quantities of interest (e.g., \ewha) 
both based on single-spaxel (\sisp) determinations and a statistics analysis of \sisp\ measurements 
within isophotal annuli (\isan).
\section{Results \label{results}}
Figure~1 illustrates on the example of the S0 galaxy NGC\ 1167 some of the quantities determined 
and discussed in G14. 
\begin{figure}
\begin{picture}(15.0,12.0)
\put(0.0,8.5){\includegraphics[width=0.3\textwidth, viewport=20 30 520 200]{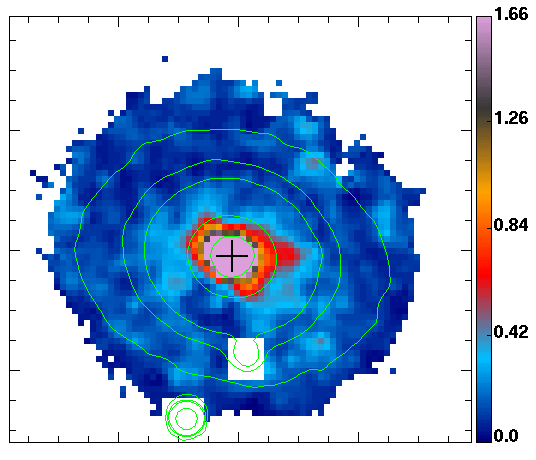}}
\put(0.0,4.5){\includegraphics[width=0.3\textwidth, viewport=20 30 520 200]{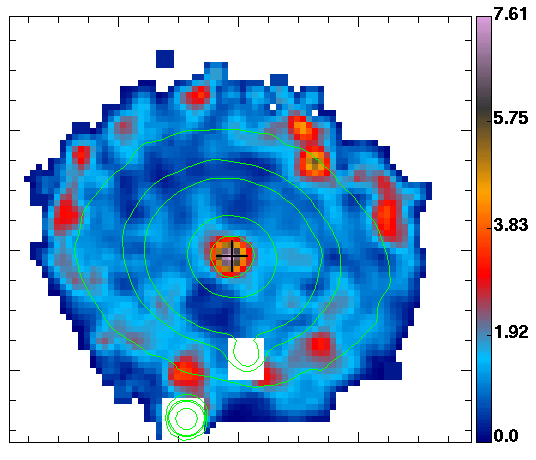}}
\put(0.0,0.5){\includegraphics[width=0.3\textwidth, viewport=20 30 520 200]{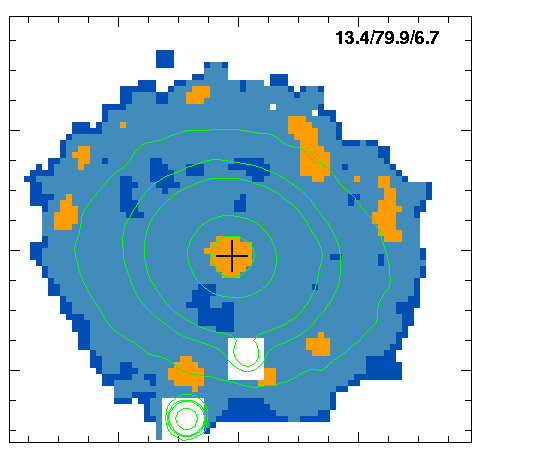}}
\put(4.0,3.7){\includegraphics[width=0.49\textwidth, viewport=20 30 420 200]{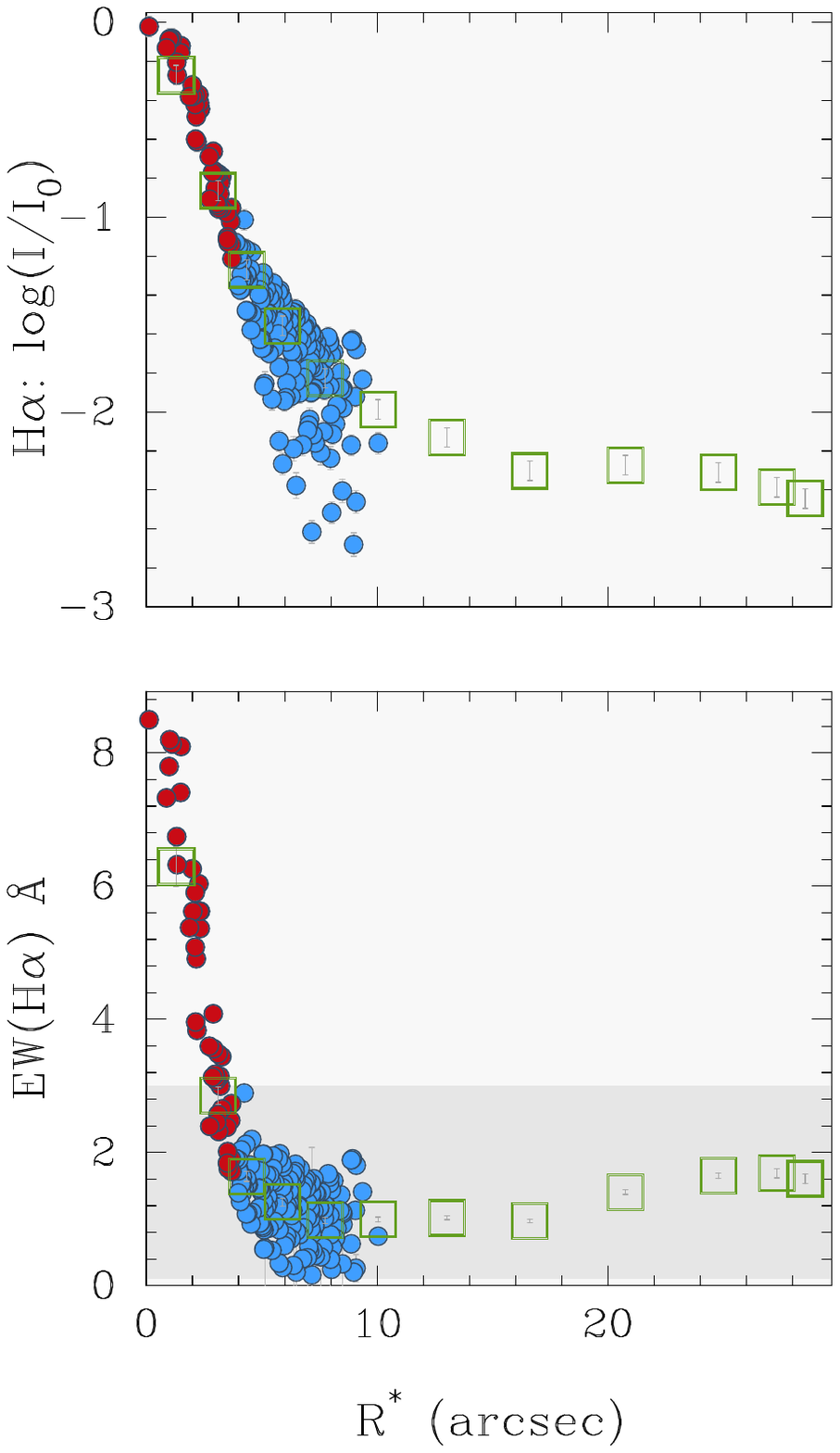}}
\put(5.0,0.35){\includegraphics[width=0.104\textwidth, viewport=20 30 420 200]{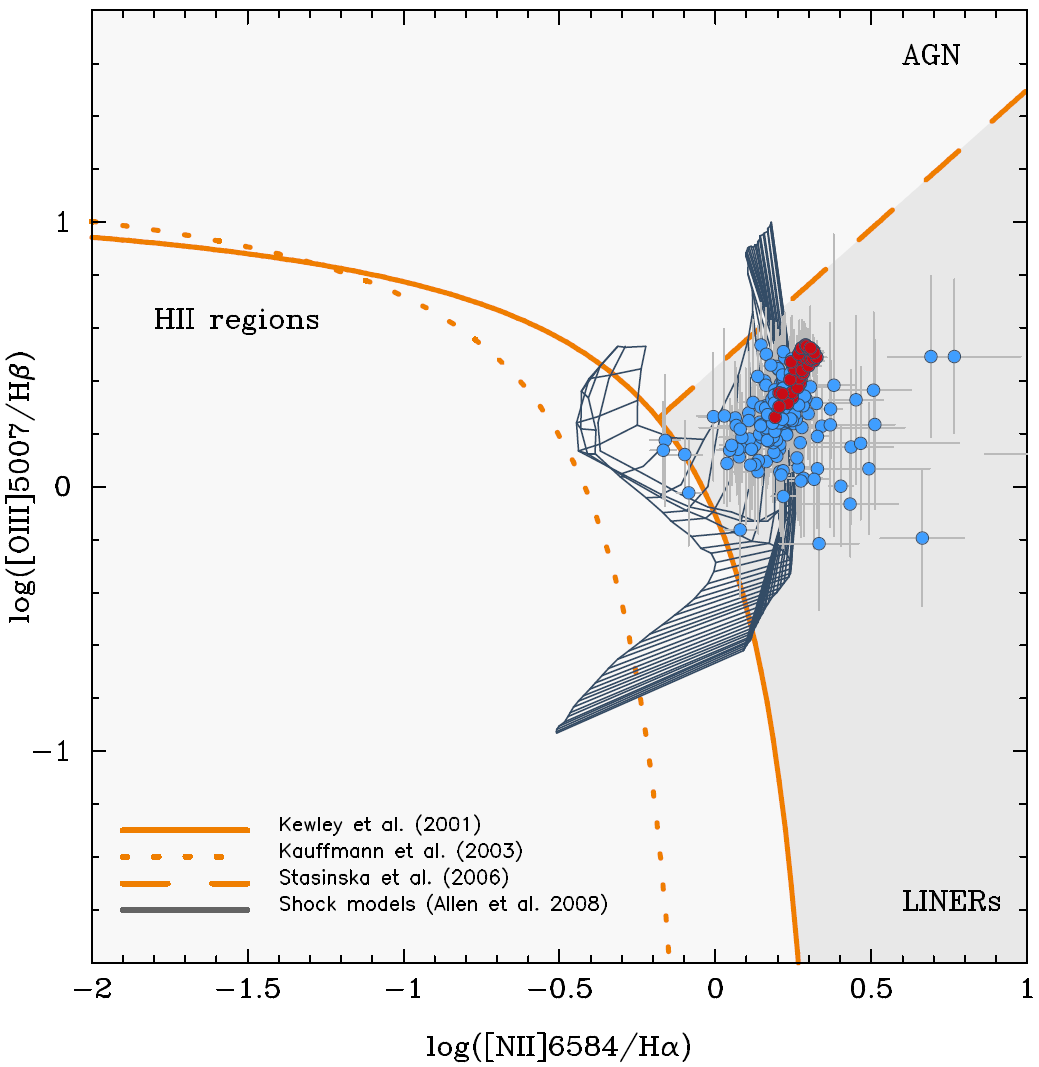}}
\put(9.5,8.7){\includegraphics[width=0.23\textwidth, viewport=20 30 420 200]{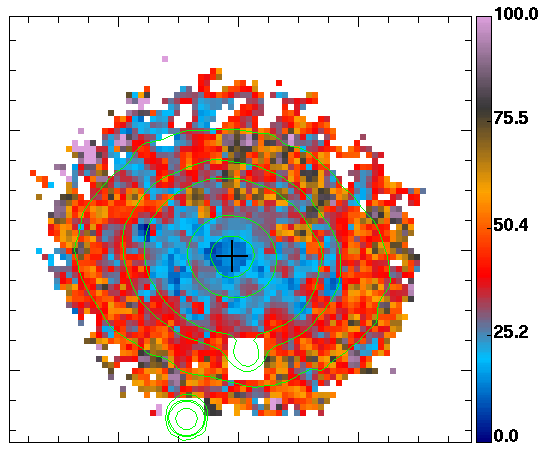}}
\put(9.5,4.5){\includegraphics[width=0.14\textwidth, viewport=20 30 420 200]{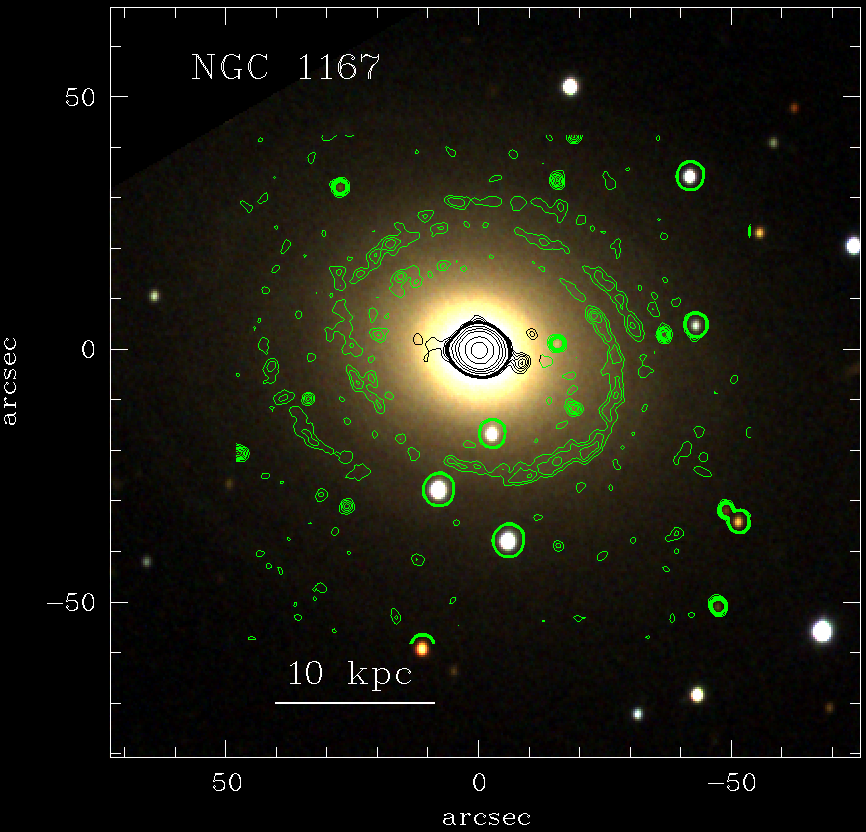}}
\put(9.4,0.4){\includegraphics[width=0.12\textwidth, viewport=20 30 420 200]{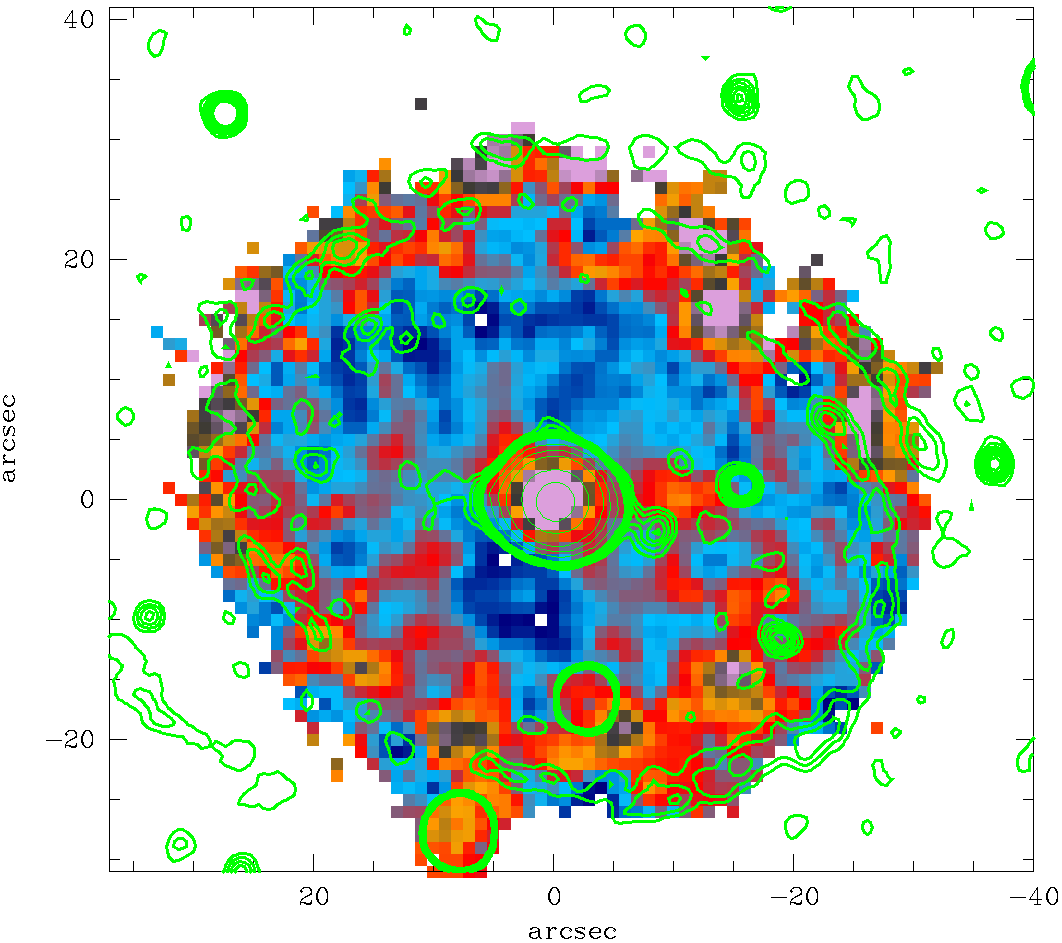}}
\PutLabel{0.1}{11.5}{\vcap a)}
\PutLabel{0.1}{7.4}{\vcap b)}
\PutLabel{0.1}{3.5}{\vcap e)}
\PutLabel{8.4}{11.5}{\vcap c)}
\PutLabel{8.4}{7.8}{\vcap d)}
\PutLabel{5.4}{3.5}{\vcap f)}
\PutLabel{12.6}{11.5}{\vcap g)}
\PutLabel{13.6}{7.6}{\vcap h)}
\PutLabel{13.0}{3.6}{\vcap i)}
\end{picture}
\caption[]{NGC\ 1167: {\trem a}--{\trem d)} \ha\ and \ewha\ maps and radial profiles; 
{\trem e)} Diagnostic subdivision of the \ewha\ map (see discussion); {\trem f)) BPT diagram;
{\trem g)} Luminosity contribution of stars younger than 5~Gyr; 
Contours overlaid with a true-color SDSS image and the \ewha\ map 
({\trem h} \& {\trem i}, respectively} depict faint spiral-like features 
in the periphery of the ETG, as revealed by image processing with an unsharp-masking technique.}
\label{fig1}
\end{figure} 
Panels \trem{a\&b} show the \ha\ flux in $10^{-16}$ \uflux\ (cf vertical bar to the
right of the panel) and \ewha\ in \AA, as determined after subtraction of the best-fitting 
synthetic stellar spectrum at each spaxel.
The radial distribution of these two quantities as a function of the photometric 
radius \rr\ (\arcsec) is shown in panels \trem{c)} and \trem{d)}, respectively. 
\sisp\ determinations obtained within the nuclear region,
defined as twice the angular resolution of the IFS data (i.e. for
\rr$\leq$3.8 arcsec) and in the extranuclear component are shown with red and
blue circles, respectively.  Open squares correspond to \isan\ determinations
within irregular isophotal annuli with the vertical bars illustrating the 
$\pm$1$\sigma$ scatter of individual \sisp\ measurements.  

A subdivision of the \ewha\ map into three intervals (panel \trem{e)}) is meant 
to help the reader to distinguish between regions where the observed \ewha\ is 
consistent, within the uncertainties, with pure pAGB photoionization (0.5--2.4~\AA; light blue), an additional gas excitation
source is needed to account for the \ewha\ ($\geq$2.4~\AA; yellow), and where
the \ewha\ is by a factor $\geq$2 lower than that predicted from pAGB
photoionization models ($\leq$0.5~\AA; dark blue), hence
\lyc\ photon escape is important.  The percentage of the spectroscopically
studied area that is consistent with these three interpretations 
is indicated at the upper-right.

Panels \rem{f)} shows \sisp\ and \isan\ determinations of the 
\tln2ha\ vs \tlo3hb\ diagnostic emission-line ratios after 
\citet[][referred to in the following as BPT ratios]{bpt81}.   
The meaning and color coding of the symbols is identical to that in panels
\trem{c\&d}. The loci on the BPT diagrams that are characteristic of AGN and
LINERs, and that corresponding to photoionization by young massive stars in
H{\sc ii} regions are indicated by demarcation lines from \citet[][dotted
  curve]{Kauffmann03}, \citet[][solid curve]{Kewley01} and \citet[][dashed
  line]{Schawinski2007}. The grid of thin-gray lines roughly at the middle of
each diagram depicts the parameter space that can be accounted for by pure
shock excitation, as predicted by \cite{Allen2008} for a magnetic field of
1~$\mu$G, and a range of shock velocities between 100 and 1000 \kmsec, for gas
densities between 0.1 and 100 cm$^{-3}$.

The luminosity contribution \lfrac\ (\%) of stars younger than 5~Gyr at
the normalization wavelength (panel \trem{g)}) echoes the
well established fact that ETGs are dominated by an evolved stellar
component throughout their optical extent. Note the clear trend for 
an outwardly increasing \lfrac, which is consistent with inside-out 
galaxy growth, or even the presence of low-level star-forming activity in the galaxy periphery.  

The overlaid contours on the true-color SDSS image (panel \trem{h}) depict  
faint spiral-like features in the periphery of NGC\ 1167, as revealed by image processing  
with the flux-conserving unsharp-masking technique by \citet[][]{P98}.
As apparent from panel \trem{i}, these contiguous low-surface brightness features 
are spatially correlated with moderately extended zones of enhanced \ewha. 
This indicates that they are not purely stellar relics from fading spiral arms that have long ceased
forming stars, but sites of ongoing low-level star formation in the extreme periphery of NGC\ 1167 
(see detailed discussion of this subject in G14).

\section*{Acknowledgements}

\noindent JMG is supported by a Post-Doctoral grant, funded by FCT/MCTES
(Portugal) and POPH/FSE (EC) and PP by an FCT Investigador 2013 Contract,
funded by FCT/MCTES (Portugal) and POPH/FSE (EC). They acknowledge support by
the Funda\c{c}\~{a}o para a Ci\^{e}ncia e a Tecnologia (FCT) under project
FCOMP-01-0124-FEDER-029170 (Reference FCT PTDC/FIS-AST/3214/2012), funded by
FCT-MEC (PIDDAC) and FEDER (COMPETE).

\end{document}